\documentclass[twocolumn]{revtex4-2}
\usepackage{physics}
\usepackage{amsmath}
\usepackage{amssymb}
\usepackage{amsfonts}
\usepackage{latexsym}
\usepackage{mathtools}
\usepackage{natbib}
\usepackage{color}
\usepackage{graphicx}
\newcommand{\beq}{\begin{equation}}
\newcommand{\eeq}{\end{equation}}
\newcommand{\half}{\frac{1}{2}}
\begin{document}
\title{Surface gravity wave on a neutron star ocean trapped around a magnetic pole}
\author{Shin'ichirou Yoshida}
\email{syoshida@g.ecc.u-tokyo.ac.jp}
\affiliation{Department of Earth Science and Astronomy, Graduate School of Arts and Sciences, The University of Tokyo\\
Komaba 3-8-1, Meguro-ku, Tokyo 153-8902, Japan}
\begin{abstract}
A warm neutron star is expected to have a fluid "ocean" of heavy elements at its outermost part of the outer crust.
As is on the terrestrial ocean, the neutron star ocean also has surface gravity waves. Around a magnetic
pole of a star, the ocean may have a dip due to the strong magnetic pressure coming into play in the hydrostatic
balance of the ocean. The dip enables the surface gravity wave to be trapped around the magnetic
pole to form eigenmodes. The frequency of the mode is much lower than the dynamical frequency at
the stellar surface, owing to the weak Coriolis force and the gradient in the ocean's depth that makes the eigenmodes present. We solve
the equation of surface gravity waves in the local $\beta$-plane approximation and obtain the spectrum of discrete eigenmodes.
We see that there are no axisymmetric modes and that the mode frequency decreases and asymptotes to zero
as the number of nodes of the corresponding
eigenfunction increases. This is reminiscent
of the g-modes in the context of asteroseismology.

Observations of X-ray binaries containing neutron stars reveal that some of the systems exhibit low-frequency
quasi-periodic oscillations (QPOs) whose frequency is $1-10^3$mHz. We investigate whether the eigenmodes
considered here may explain the low-frequency QPO spectrum. It is suggested that some of the QPOs in the system whose neutron star spins at the period less than $10$s may be consistent with the model. As far as the
spin period is larger than $10$s, the eigenmode frequencies are too low to explain the observed QPOs.
\end{abstract}
\maketitle

\section{Introduction}
The surface layer of neutron stars is usually assumed to be a solid crystal made of heavy metal elements. When a star is warm enough ($T\gtrsim 10^6$K), a fluid "ocean" covers the solid crust.
According to \cite{Urpin2004}, if the atomic number $Z$
of the surface element is below $26$, the crystalization occurs at the density $\rho\ge 10^9\mbox{g}\mbox{cm}^{-3}$,
which leads to a fluid layer depth of $\sim 100$m. Therefore, the ratio of the ocean's depth to the typical radius of 
a neutron star is ${\cal O}[10^{-2}]$ (for the terrestrial ocean, the ratio is ${\cal O}[10^{-3}]$.).
This may happen when it retains its latent heat at its birth or when it is heated by the accretion
and thermonuclear reactions of the gas supplied from its binary companion. 
Weather-like dynamics of a neutron star ocean and atmosphere are discussed in connection to the vortices and zonal jets reminscent to those observed on Earth, which may affect the interpretations of observations
of X-ray binaries \cite{Nattila_etal2024}.
On the other hand, a fluid layer under gravitational acceleration supports classes of oscillations, and they have been discussed
in relation to X-ray observations of accreting stars. \cite{Bildsten_Cutler1995, Bildsten_etal1996} studied a model that explains
the quasi-periodic oscillations (QPO) of X-ray flux in low-mass X-ray binaries (LMXBs) with frequencies of a few Hz by the
global g-mode oscillations of the ocean. \cite{Heyl2004} discussed that another type of global modes (r-modes), which have the Coriolis
force as a restoring force, may be relevant to the QPOs. \cite{vanBaal_etal2020} investigated the effect of rapid rotation of
the star on the global oscillation modes (Poincar\'{e}-, Yanai- or r-mode) of the thin ocean, by taking into account the oblate deformation of the star due to the rotation. The oscillations that the latter authors studied have the surface deformation of the ocean as a restoring mechanism, and are regarded
as global eigenmodes of the surface gravity waves.
At the boundary of the layer with another fluid (or atmospheric) layer or with a vacuum,  the deformation of the surface
travels at a phase velocity depending on the depth of the ocean. This surface gravity wave has long been studied in
the context of geophysics, especially of the tidal waves on the terrestrial ocean. Tsunamis excited at a large earthquake 
may travel on the ocean to reach the coasts of different continents. The analyses of them are usually done by assuming the
shallow water approximation, i.e., the horizontal length scale is much larger than the depth of the sea.
Unless there is a 'wall' surrounding the propagation region, the surface gravity wave is not trapped and it exists as a traveling wave
\footnote{The Atlantic Ocean is regarded as such a case with walls. See, M. Rieutord, arXiv:astro-ph/0308313 [astro-ph].} .
The nature of the wave, however, may change when the gradient of the depth of the ocean is taken into account.
Since the phase velocity of the surface gravity waves depends on the local depth of the sea, the waves may
be diffracted when the depth has a horizontal gradient. 
There have been claims \cite{Chambers1965, Longuet-Higgins1967, Longuet-Higgins1969} that a coast around an island 
which has a shallower seabed nearer to it, may diffract the tidal wave to form a trapped oscillation around
the island. This is possible because the phase velocity of the wave is $\sim \sqrt{gh}$, where $g$ is the gravitational
acceleration and $h$ is the depth of the sea. From the Huygens principle, the traveling wave may circle around
the island to form a quantized state of the sea level. It has been claimed by some geophysicists that these trapped modes
were observed as aftermaths of tsunamis caused by some earthquakes (e.g, \cite{Fujima_etal1995}).

A neutron star has a strong magnetic field that may affect its ocean. The surface of the ocean at a magnetic pole
may be deformed by Maxwell's stress so that the surface gravity waves form bound states with characteristic frequencies.
This may be relevant to some quasi-periodic oscillations (QPOs) observed in the accreting X-ray binaries. If the magnetic field 
is high enough, the accreting gas forms a column at a magnetic pole, which emits X-rays. The modulation of X-ray other than
the stellar spin frequency has been found, and its relation to the oscillations of the accretion column is discussed.
If there are trapped oscillation modes of the ocean at the base of the magnetic pole, some of these QPOs may be
related to the excitation of these modes. 

In this paper, we introduce a model of the surface gravity wave around the magnetic pole and show
that the discrete eigenmodes are possible. There are no axisymmetric eigenmodes, 
and the spectral characteristics of eigenmodes are reminiscent of the g-modes of a stably stratified star, though the physical
mechanism is different.

\section{Formulation}

\subsection{Assumptions}
For simplicity, we consider the neutron star ocean to be composed of incompressible fluid, since the surface gravity wave of
our interest is subsonic. Newtonian mechanics is employed with relativistic corrections to be taken into account when the eigenfrequencies are evaluated. We assume the magnetic field is axisymmetric around a magnetic pole (not necessarily one of the two poles of the conventional dipole field model of a neutron star. The field may be of one of the local magnetic poles for a multipolar or disordered field configuration).
Since our interest is the local wave propagation or trapping, we introduce a local Cartesian coordinate whose origin
sits at the intersection of the magnetic axis and the bottom of the sea.  The magnetic field is assumed to be confined in a finite cylindrical volume, whose radius is $r_s$. The curvature effect of the stellar surface is thus neglected. The local gravity is uniform and anti-parallel to the local z-axis. 

We assume the bottom of the sea is uniform in the region of interest and has a planar geometry. The z-coordinate is chosen to be normal to the bottom of the sea and is directed outward. 

If the seabed were not on the equipotential surface, the boundary condition of the surface gravity wave 
at the bottom of the sea would not be as simple as is adopted here, and the variable seabed level should be 
taken into account in the wave equation. This might happen around the magnetic pole, where the strong field
modified the crystal-fluid phase transition of the crust matter. In fact, we see it is not necessary to introduce 
this complication to the model.
While it is generally true that a strong magnetic field may influence the crust geometry, the studies on magnetic
crust matter show that the magnetic field has a negligible effect to determine the fluid-solid boundary (seabed) under the physical conditions considered in this study. First, it has been shown that the melting temperature of the crust as a function of density does not significantly depend on the magnetic field strength. If it did, the seabed would not lie on an equipotential surface. For the density region of interest ($\rho \le 10^9 \text{ g/cm}^3$; \cite{Urpin2004}), the melting temperature changes very little for a typical pulsar's field strength of $B = 10^{12} \text{ G}$ (see, e.g., Fig. 2 in \cite{Chamel-Haensel2008}). Furthermore, for an accreting neutron star with a near-Eddington accretion rate ($\dot{M} = 10^{-10}$ to $10^{-8} \, M_\odot/\text{yr}$), recent simulations by \cite{Nava-Callejas2025} show that the relevant density domain maintains a high temperature of several $\times 10^7$ to several $\times 10^8 \text{ K}$. Therefore, the magnetic field is not expected to affect the fluid-solid phase transition. Second, the strongly magnetized liquid atoms may be susceptible to "magnetic condensation" \cite{Chamel-Haensel2008}), where strong magnetic fields confine electrons to Landau levels and deform atoms. If the magnetic condensation of atoms occured, the elongated
atoms would align with the magnetic field and form a crystal-like structure, and the magnetic field would affect
the position of the seabed. According to \cite{Medin_Lai2007}, the critical temperature $T_c$ for $^{56}\text{Fe}$ below which this phase appears is $T_c = 7 \times 10^5 \text{ K}$ for $B = 10^{13} \text{ G}$, $T_c = 3 \times 10^6 \text{ K}$ for $B = 10^{14} \text{ G}$, and $T_c = 2 \times 10^7 \text{ K}$ for $B = 10^{15} \text{ G}$. Since the temperature of an accreting neutron star is significantly higher than these critical temperatures, magnetic condensation is unlikely to occur in our targeted system. Consequently, we can safely assume that the seabed lies on the equipotential surface of the slowly rotating star
regardless of the magnetic field, and the equilibrium ocean depth is determined by the magnetohydrostatic balance above it.

It should be also noted that the magnetic pressure at the seafloor is three orders of magnitude smaller than the gas pressure, 
if we adopt the typical field strength of $B\sim 10^{12}$G (e.g., Fig.2.2 of \cite{Shapiro-Teukolsky}). 
It follows that the density stratifcation in the crust at which the liquid-solid phase transition takes place, 
is the same as if the magnetic field was absent. Thus the seabed is on the equipotential surface of the star.
The magnetic contribution to the hydrostatic balance becomes larger for the shallower point in the ocean.
We therefore take into account the magnetic pressure to determined the depth of the ocean. 

Since the local coordinate rotates at the stellar angular 
frequency $\Omega_\star$, there appear inertial force terms in the equation of motion. We neglect the centrifugal term by assuming 
$\Omega_\star$ is small compared to the Keplerian break-up limit of the star. Then the Coriolis force term is added to the tangential components
of the equation of motion. Their components in x- and y-direction are respectively $-fv^y$ and $fv^x$, where
the Coriolis factor $f$ is defined by $f = 2\Omega_\star\cos\theta$, by using the positional polar angle $\theta$ of the magnetic pole with respect
to the rotational axis of the star. This is the lowest order $\beta$-plane approximation broadly adopted in the context of meteorology or oceanology of the Earth.

The magnetic field is parallel to the z-axis (its direction, upward or downward, is not significant in the current model)
and axisymmetric with respect to the z-axis, $\mathbf{B} = B^z(r) \mathbf{e}_z$, where $\mathbf{e}_z$ is the unit normal vector
in z-direction, and $r=\sqrt{x^2 + y^2}$ is the cylindrical radius.

\subsection{Equilibrium profile of the fluid pressure, the magnetic field and the surface profile of the ocean}
We assume that the equilibrium state of the shallow ocean is static and axisymmetric with respect to the magnetic
pole. We have
\beq
	\frac{1}{\rho}\pdv{p}{x} + \frac{1}{8\pi\rho}\pdv{B^2}{x} = 0,
	\label{eq: equil force x}
\eeq
\beq
	\frac{1}{\rho}\pdv{p}{y} + \frac{1}{8\pi\rho}\pdv{B^2}{y} = 0,
	\label{eq: equil force y}
\eeq
\beq
	\frac{1}{\rho}\pdv{p}{z} + g = 0.
	\label{eq: equil force z}
\eeq
Here $p(x, y)$ is the fluid pressure, $B(x, y)$ is the vertical magnetic field strength and $g$ is the gravitational
acceleration. From (\ref{eq: equil force x}) and (\ref{eq: equil force y}) we have
\beq
	\frac{p}{\rho} + \frac{B^2}{8\pi\rho} \equiv \Pi(z),
\eeq
where $\Pi$ is a function of $z$. Then Eq.(\ref{eq: equil force z}) is integrated as
\beq
	\Pi(z) = -gz + C,
\eeq
where $C$ is a constant. We assume that the magnetic field is nonzero if $\sqrt{x^2 + y^2}\le r_s$ for some
radius $r_s$. Then the exterior region to the magnetic pole has no magnetic field, and the height of the
surface of the ocean is constant $h(x, y)=h_{out}$ there. Evaluating $C$ at the seabed outside the magetic
pole, we have $C = p_b/\rho$, where $p_b$ is the pressure at the seabed. Then we have
\beq
	\frac{p_s^{int}}{\rho} + \frac{B^2}{8\pi\rho} + g\bar{h}(x, y) = \frac{p_s^{ext}}{\rho} + g h_{out} = \frac{p_b}{\rho}.
\eeq
Here $p_s^{int}$ and $p_s^{ext}$ are pressure at the surface of the ocean, i.e., the pressure of the accreted matter
(atmosphere of the neutron star), interior and exterior to the magneticl pole region. We may assume $p_b\gg p_s^{int}, p_s^{ext}$ and neglect the surface pressure
terms. Then we have,
\beq
	\frac{B^2(x, y)}{8\pi\rho} = g(h_{out} - \bar{h}(x, y)).
\eeq
This is seen as a definition of $\bar{h}$ when the profile of $B(x, y)$ is known. Instead, we may regard
it as the equation defining the magnetic field profile when the surface profile of the ocean is given.
In this paper, we adopt the latter viewpoint and give a simple analytic profile of $\bar{h}$.
The fluid pressure $p(x, y, z)$ is then determined by,
\beq
	\frac{p(x, y, z)}{\rho} = g(\bar{h}(x, y) - z).
\eeq
By the assumption of axial symmetry, $B, \bar{h}$ are functions of cylindrical radial coordinate $r$, while $p$
is  the function of $(r, z)$.

Suppose the difference between $h_{out}$ and $h(r)$ is $\sim (1-\alpha) h_{out}~(0<\alpha<1)$. The scale of the magnetic field
is evaluated as,
\begin{widetext}
\beq
	B \sim \qty[8\pi\rho g(1-\alpha) h_{out}]^\half \sim 2\times 10^{15}{\mbox (G)}~
	(1-\alpha)^\half\qty(\frac{M}{1.4M_{\odot}})^\half \qty(\frac{R}{10{\mbox km}})^\half
	\qty(\frac{\rho}{10^9{\mbox g}{\mbox cm}^{-3}})^\half \qty(\frac{h_{out}}{R})^\half
	\label{eq: B-field estimation}
\eeq
\end{widetext}
where $M$ is the stellar mass, $R$ is its radius, $\rho$ is the mass density of the ocean.

\subsection{Linearized dynamical equation}
Time evolutions of fluid velocity $v$ and magnetic field $B$ are determined by
\begin{align}
	\pdv{v_a}{t} + v^b\grad_b v_a = -\frac{1}{\rho}\grad_a p - g_a + \frac{1}{c\rho}\epsilon_{abe}j^bB^e,\label{eq: momentum}\\
	\pdv{B^a}{t}  = \epsilon^{abd}\grad_b(\epsilon_{def}v^eB^f),\label{eq: induction}
\end{align}
where the first equation is the fluid equation of motion with Lorentz force, and the second is the magnetic induction equation
for the ideal magnetohydrodynamics (MHD) approximation (infinite conductivity, and time variation of the electric field is neglected).
Here $\epsilon_{abd} = \sqrt{\mathfrak{g}}[a,b,d]$, $\mathfrak{g}$ is the determinant of the 3-metric coefficients $\mathfrak{g}_{ab}$
and $[a,b,d]=\pm 1$ for even/odd permulation of $1,2,3$.
Notice that $\epsilon^{abd} = \frac{1}{\sqrt{\mathfrak{g}}}[a,b,d]$.
$g_a$ is the component of the gravitational acceleration vector.

The electric current density $j^a$ is related to the field by the Amp\`{e}re-Maxwell's equation
\beq
	\epsilon^{abd}\grad_bB_d = \frac{4\pi}{c}\qty(j^a+\pdv{E^a}{t}).
	\label{eq: Ampere-Maxwell}
\eeq
In the standard MHD approximation, the displacement current term is neglected when the characteristic
velocity of the system is smaller than the speed of light. We see that the low-frequency modes considered here
satisfy this condition. We thus omit the displacement current term and use Eq.(\ref{eq: Ampere-Maxwell})
to express the current density in terms of the magnetic field
	\footnote{Assuming the stellar radius of $10$km and the mass of $1.4M_\odot$, we see the phase velocity
	of the surface gravity wave in the shallow-water approximation, $\sqrt{GM/R^2 h}$, is the order of a few percent of the speed of
	light, when the depth of the ocean is $10^2$m.}.
The third term on the right-hand side of Eq.({\ref{eq: momentum}}) is written as
\begin{align}
	\frac{1}{4\pi\rho}\qty(B^b\grad_bB_a - B^b\grad_aB_b) = &&-\frac{1}{4\pi\rho}\grad_a(\frac{1}{2}B^bB_b) \nonumber\\
	&&+ \frac{1}{4\pi\rho}B^c\grad_c B_a
\end{align}

We work in the linearized version of these equations. The Eulerian perturbation of quantity $q$ is denoted as $\delta q$.
Assuming the shallow water approximation detailed in the Appendix, we have the perturbed set of equations as follows.
The tangential  components of the equation of motion are
\beq
	\pdv{\delta v^x}{t} = -f\delta v^y - g\pdv{\eta}{x},
\eeq
and
\beq
	\pdv{\delta v^y}{t} = f\delta v^x - g\pdv{\eta}{y}.
\eeq

The continuity equation is expressed by the perturbed incremental height of the ocean surface $\eta(x, y, t)$ as
\beq
	\pdv{\eta}{t} + \pdv{x}\qty[\bar{h}\delta v^x] + \pdv{x}\qty[\bar{h}\delta v^x] = 0.
\eeq
Here $\bar{h}(x, y)$ is the equilibrium height of the surface with respect to the sea floor.
The coefficients of the equations do not depend on $t$, thus we have the temporal harmonic
decomposition of the perturbed quantities as $\eta \propto e^{-i\sigma t}$. Therefore, we can cast the
equations above to the second-order partial differential equation for $\eta$,
\begin{widetext}
\beq
	\sigma\eta + \frac{\sigma}{\sigma^2-f^2}\qty[\pdv{x}\qty(\bar{h}\pdv{\eta}{x})
	+ \pdv{y}\qty(\bar{h}\pdv{\eta}{y})]
	+ \frac{if g}{\sigma^2-f^2}\qty[\pdv{x}\qty(\bar{h}\pdv{\eta}{y}) - \pdv{y}\qty(\bar{h}\pdv{\eta}{x})]
	= 0.
\eeq
\end{widetext}
Now we transform the coordinate to the cylindrical one, $(r, \phi, z)$.
Since the system is axisymmetric, we may separate the $\phi$ dependence of $\eta$ as
$\eta\propto e^{im\phi}$ and the equation is cast into the master equation,
\begin{widetext}
\beq
	\dv[2]{\eta}{r} + \qty(\frac{1}{r} + \dv{r}\ln\bar{h})\dv{\eta}{r}
	+ \qty[-\frac{m^2}{r^2} + \frac{\sigma^2-f^2}{g\bar{h}} + \frac{mf}{r\sigma}\dv{r}\ln\bar{h}]\eta = 0.
	\label{eq: master equation}
\eeq
\end{widetext}

\subsection{Boundary conditions}
Eq.(\ref{eq: master equation}) is supplemented with appropriate boundary conditions to form an eigenvalue problem for $\sigma$.
At $r=0$, $\eta$ must be regular, which leads to the condition that $\eta\sim r^m$ as $r\to 0$.
As for the outer boundary condition, we are interested in the trapped oscillations, which means
that the eigenfunction damps exponentially at $r\to\infty$.
We note that the magnetic field is confined inside a finite radius $r_s$. For $r\ge r_s$, we have $B^z=0$, therefore $\dv{\bar{h}}{r}=0$. In this outer region, Eq.(\ref{eq: master equation}) reduces to
the modified Bessel's eqution
\beq
	\dv[2]{\eta}{r} + \frac{1}{r}\dv{\eta}{r}
	+ \qty[-\frac{m^2}{r^2} + \frac{\sigma^2-f^2}{g\bar{h}}]\eta = 0,
	\label{eq: equation in the outer region}
\eeq
where $\bar{h}$ is constant. Eq.(\ref{eq: equation in the outer region}) has two linearly independent solutions,
one of which is exponentially growing as $r\to\infty$, thus should be discarded. Then, we have the outer boundary condition
$\eta(r)\propto K_m(r\sqrt{(f^2-\sigma^2)/g\bar{h}})$ for $r\ge r_s$. Here $K_m(x)$ is the modified 
Bessel's function of the second kind. It is worth commenting that eigenmodes whose eigenfunction
localizes around the magnetic pole
when $\sigma^2\le f^2$, i.e., in the low frequency range compared with the stellar rotational
frequency. Otherwise, Eq.(\ref{eq: equation in the outer region}) is instead cast into Bessel's equation
with the integer order $m$.  Linearly independent 
solutions of it behave as $J_m(r\sqrt{(\sigma^2-f^2)/g\bar{h}})$ or $Y_m(r\sqrt{(\sigma^2-f^2)/g\bar{h}})$,
where $J_m(x)$ and $Y_m(x)$ are Bessel's functions of the first and second kind. Neither of these solutions damps 
exponentially as $r\to\infty$. Rather, they represent traveling surface waves, which are scattered
by the magnetic pole.

\section{Results}
\begin{figure}[h]
\includegraphics[width=9cm]{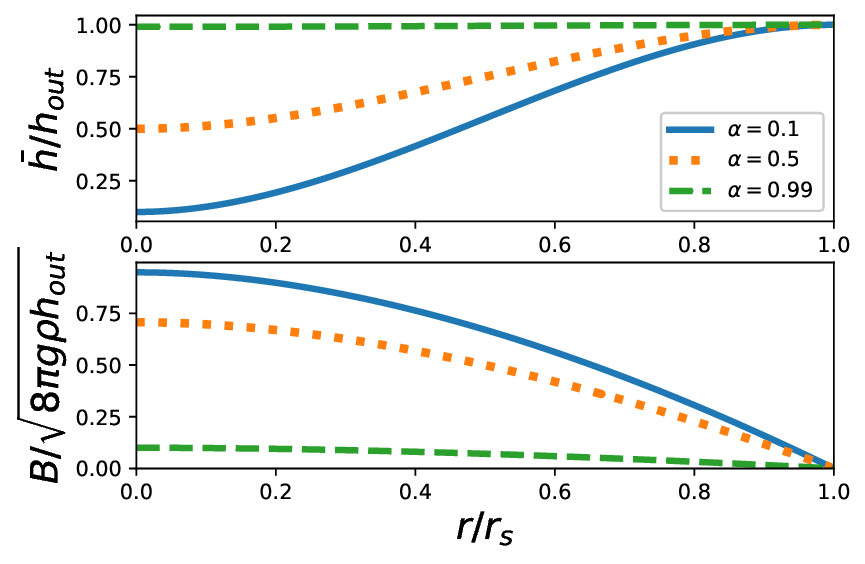}
\caption{\label{fig: profile of h}  (Top) Dimensionless profile of $\bar{h}(r)$. The distance from the magnetic axis
$r$ is normalized by the radius of the magnetic pole $r_s$. The depth $\bar{h}(r)$ is normalized by the
depth $h_{out}$ outside the magnetic pole.
(Bottom) Dimensionless profile of the magnetic field.}
\end{figure}

For simplicity, we choose the equilibrium profile of depth $\bar{h}(r)$ as an analytic function (see 
the top panel of Fig.\ref{fig: profile of h}):
\beq
\bar{h}(r) =
\begin{cases}
	6h_{out} (1-\alpha) \qty(\frac{r^2}{2r_s^2} - \frac{r^3}{3r_s^3}) + \alpha h_{out} &  r\le r_s\\
	h_{out} & r\ge r_s.
\end{cases}
\eeq

Here $h_{out}$ is the constant depth outside the magnetic pole. For dimensionless expression of $h_{out}$, $\gamma = r_s/h_{out}$ is introduced 
as the ratio of the radius of the magnetic pole $r_s$
to the average depth $h_{out}$. 
The parameter $\alpha$ ($0\le\alpha\le 1$) controls the relative change of the depth due to the magnetic field. The larger $\alpha$ corresponds
to the smaller depth change inside the magnetized region. The first derivative of the depth
vanishes both at $r=0$ and $r=r_s$, ensuring that the surface profile of the ocean is smooth everywhere.
The equivalent definition of equilibrium magnetic field is shown in the bottom panel.
For the current choice, the magnetic field monotonically decreases as a function of the cylindrical radial
coordinate. The magnetic profile leads to the magnetic pressure gradient pointing inward, which
pushes the fluid outward. This force is balanced by the pressure gradient that points outward, which confines the magnetic field inside the pole. 

\subsection{Characteristics of the eigenvalue problem}
First, we look at the dispersion relation of the surface gravity wave. By transforming the dependent function
as $\xi\equiv(r\bar{h})^{1/2}\eta$, Eq.(\ref{eq: master equation}) is cast into a Schr\"{o}dinger-like
form as,
\beq
	\frac{d^2\xi}{dr^2} + \left[-\frac{m^2}{r^2} + \frac{1}{\sigma}\frac{mf}{r^2}\frac{d\ln\bar{h}}{d\ln r}\right]\xi\approx 0,
\eeq
where the dominant terms with low frequency mode $|\sigma|<<1$ are retained.
Then, we have the dispersion relation by assuming $\xi\propto e^{ikr}$ as,
\beq
	\sigma \approx \frac{mf}{m^2+k^2r^2}\left(\frac{d\ln\bar{h}}{d\ln r}\right).
\eeq
The frequency of the wave is, therefore, determined by the Coriolis parameter $f$ (proportional to the rotational frequency
of the star) as well as the gradient of the equilibrium ocean's depth.

Next, we see the characteristics of the eigenvalue problem of Eq.(\ref{eq: master equation}) with
the boundary condition at $r=0$ (regularity) and $r\to\infty$ (exponentially damping).

First, we notice there are no axisymmetric trapped modes with $m=0$. To see this, we cast Eq.(\ref{eq: master equation}) with $m=0$ in the following
form,
\beq
	\dv{x}\qty(a(r)\dv{\eta}{r}) - \kappa b(r)\eta = 0,
	\label{eq: m=0 SL form}
\eeq
where $a(r) = r\bar{h(r)}$, $b(r) = (g\bar{h})^{-1}$, and $\kappa = f^2 - \sigma^2$.
This form of equation with the regular boundary conditions at $r=0$ and $r=\infty$
forms the Strum-Liouville type of eigenvalue problem (e.g., \cite{r_courant_methods_1965}).
As is seen above, a trapped mode must have $\kappa>0$. We also notice $a(r)\ge 0$
and $b(r)>0$. Integrate Eq.(\ref{eq: m=0 SL form}) from $r=0$ to $r=\infty$, we have,
\beq
	\int_0^\infty \dv{r}\qty(a(r)\dv{\eta}{r}) dr = \qty[a(r)\dv{\eta}{r}]_0^\infty = \kappa\int_0^\infty b(r)\eta dr,
\eeq
thus,
\beq
	\lim_{r\to\infty} a\dv{\eta}{r} = \kappa\int_0^\infty b(r)\eta(r) dr.
	\label{eq: integral m=0}
\eeq
The eigenfunction of the fundamental mode of the problem would have no node
when $\eta$ has a definite sign. Suppose the sign is positive. The right-hand side of
Eq.(\ref{eq: integral m=0}) is positive. Then we should have $d\eta/dr(\infty)>0$,
though it must be zero. Exactly in the same way, a negative-definite $\eta$ is also
impossible. Consequently, we have no fundamental mode without a node of its eigenfunction,
and we have no solutions to the Sturm-Liouville problem.

Next, we consider the low frequency limit, $|\sigma|\ll 1$ for fixed $m$. We find no unstable eigenmode for the equilibrium
model considered. Eq.(\ref{eq: master equation}) is cast in the standard form of a Strum-Liouville type equation,
\beq
	\dv{r}\qty(P(r)\dv{\eta}{r}) - Q(r)\eta + sR(r)\eta = 0,
\eeq
where $s\equiv m/\sigma$ and,
\begin{align}
	P(r) = r\bar{h}\ge 0,\\
	Q(r) = \qty(\frac{m^2}{r^2} + \frac{4f^2}{\bar{h}})r\bar{h}\ge 0,\\
	R(r) = 2\Omega_\star\dv{\bar{h}}{r} \ge 0.
\end{align}
The theory of Strum-Liouville type equations tells us that the eigenvalue $s$ is positive and discrete
(therefore no unstable eigenmode exists).
For a fixed value of $m$, the eigenvalues $s_n~(n=0,1,2, \dots)$ form an ascending sereies,
$0<s_0<s_1<s_2\dots$. Therefore, we have $0< \dots < \sigma_2/m < \sigma_1/m < \sigma_0/m$.
The corresponding eigenfunction $\eta$ has n nodes for the n-th eigenvalue (the eigenfunction of
the 0-th eigenmode, the fundamental mode, has no node).
We also see that the eigenvalue $\sigma$ for the azimuthal order $-m (m>0)$ is
the sign-reversed value of the case with the order $m$.
In the low frequency limit, we see the discrete eigenvalues, $\sigma/m$, accumulate to zero from above.
This nature is reminiscent of the g-modes in the asteroseismology \cite{Unno-et-al1989}.
The g-mode eigenvector is dominated by the tangential component of the velocity, which is also
the case for the surface gravity wave. This resemblance is expected, since both oscillations
share the gravity (in the form of buoyancy for g-moes) as a restoring force.

It should be mentioned that the freely-travelling surface gravity wave trapped
by some walls or obstacles would have a much higher frequency. The frequency $\nu$ of the surface gravity wave
traveling back and forth in a length scale $L$ is, $\nu = \sqrt{gh}/(2L) = 7~\mbox{kHz} \times (M/1.4M_\odot)^{1/2}(R/10\mbox{km})^{-1}(h/100\mbox{m})^{1/2}(L/1\mbox{km})^{-1}$. On the other hand, we see from Eq.(\ref{eq: master equation}) that the characteristic frequency 
of the trapped surface gravity wave scales as
\beq
	\sigma \sim f \frac{d\ln r}{d\ln\bar{h}}.
	\label{eq: scaling of frequency}
\eeq
Thus, the frequency scales with the stellar rotational frequency ($\propto f$) and the logarithmic gradient
of the equilibrium depth of the ocean.
It is the combination of the rotation of the star and the gradient of the ocean's depth produced by the magnetic pole,
that makes the existence of the low-frequency trapped eigenmode possible.

\subsection{Numerical results}
First, we show the solutions of the eigenvalue problem in the dimensionless expression.
We note that the dimensionless parameters of the problem are $\alpha$, which
measures the relative depth of the magnetized and the unmagnetized region (Fig.\ref{fig: profile of h})
and the dimensionless Coriolis factor $\tilde{f}$,
\beq
	\tilde{f} = 2\gamma\sqrt{\frac{h_{out}}{g}}\Omega_\star\cos\theta.
\eeq
Gravitational acceleration at the stellar surface, $g$, contains the relativistic correction
as mentioned in Sec.\ref{sec: discussion}. We see that $\gamma$, $h_{out}$, and $\Omega_\star\cos\theta$
are degenerate for a value of $\tilde{f}$.

\begin{figure}[h]
\includegraphics[width=9cm]{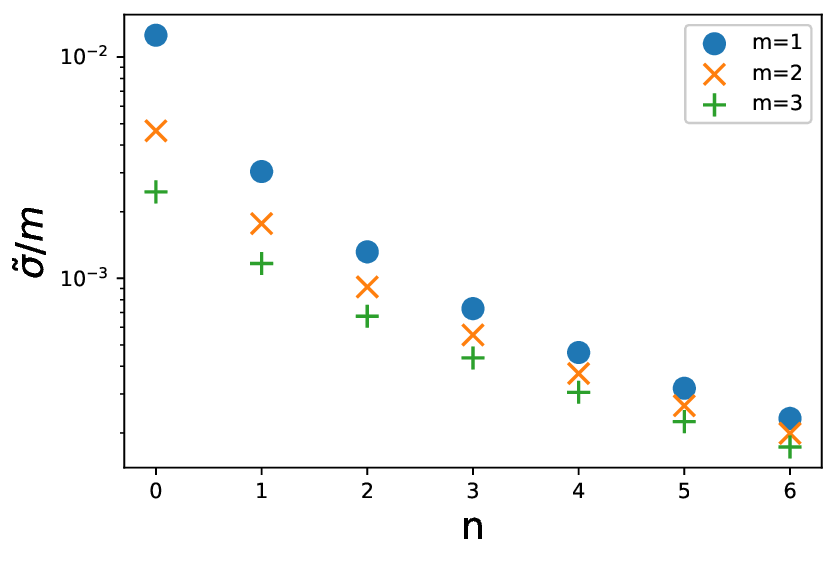}
\caption{\label{fig: spectrum} Eigenspectrum of trapped surface gravity wave. Eigenfrequency $\sigma$
is normalized by the half of the Coriolis factor $f$. $m$ is the order of the harmonic dependence
$e^{im\phi}$. The vertical axis corresponds to the pattern speed of the mode in the azimuthal direction.
The horizontal axis is the number of nodes of the eigenfunction.
The model parameters are $(\alpha, \tilde{f}) = (0.98, 0.0807)$.
The value of $\tilde{f}$ corresponds to $\gamma=10$, $\nu_{rot}\cos\theta=100$Hz, $h_{out}=100$m,
in a physical scale. }
\end{figure}

In Fig.\ref{fig: spectrum}, a part of the lowest order (thus the highest frequency) eigenspectrum is presented.
The frequency divided by $m$ is the pattern speed of the eigenmode in the azimuthal direction. We see that
the mode pattern rotates in the prograde direction with the local Coriolis parameter (thus in the same direction
as the stellar rotation). The eigenmodes with the negative $m$ have approximately the same pattern speed as
the modes with $-m$. 

\begin{figure}[h]
\includegraphics[width=9cm]{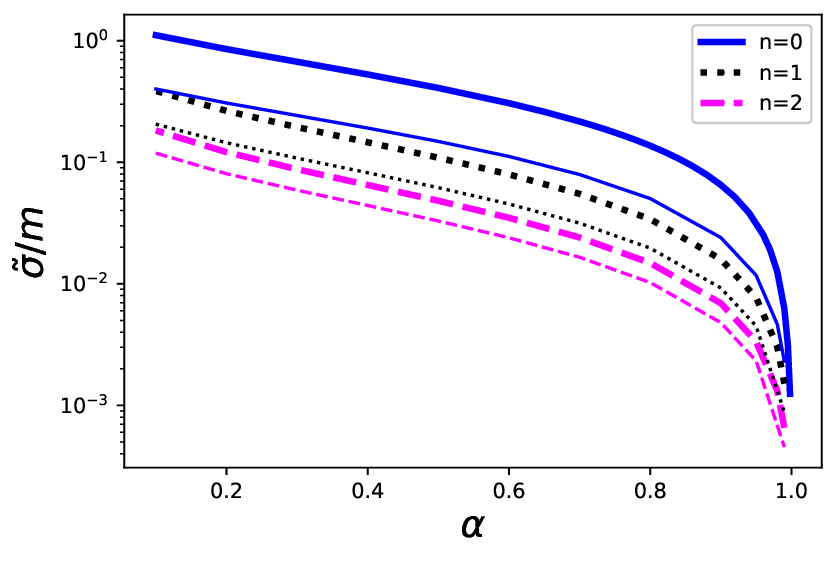}
\caption{\label{fig: eigenvalue_alpha_dependence}  Eigenvalues as a function of $\alpha$. The model parameter is $\tilde{f}=0.0807$.
The thick curves correspond to $m=1$ cases for different $n$, while the thin ones are for $m=2$.}
\end{figure}

By fixing the value of $\tilde{f}$, we examine the dependence on $\alpha$ of eigenvalues for selected modes.
Fig.\ref{fig: eigenvalue_alpha_dependence} shows three of the lowest order eigenmodes for $m=1$.
The eigenfrequency decreases as $\alpha$ increases. The smaller difference in the depth of the magnetized region
compared with $h_{out}$, or a weaker magnetic field, reduces the eigenfrequency. 
It should be noted that the limit of $\alpha\to 1-0$ ($\tilde{\sigma}\to 0$) is a singular limit of the problem. 
If we set $\alpha=1$ (no magnetic field), Eq.(\ref{eq: master equation}) reduces to  Eq.(\ref{eq: equation in the outer region}),
and there are no trapped eigenmodes. 

\begin{figure}[h]
\includegraphics[width=9cm]{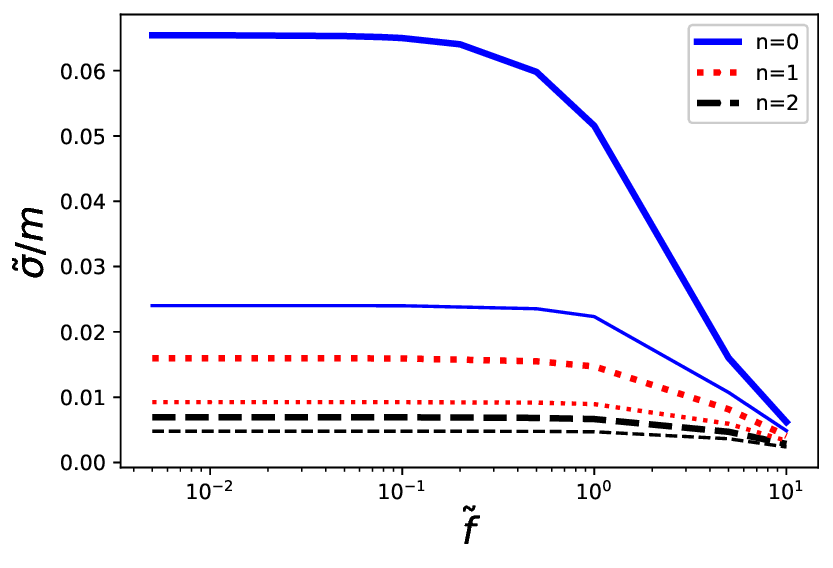}
\caption{\label{fig: eigenvalue_f_dependence} Eigenvalues as a function of $\tilde{f}$.
The model parameter is $\alpha=0.9$. The thick curves correspond to $m=1$ for different $n$, while the thin
ones are for $m=2$.}
\end{figure}

In Fig.\ref{fig: eigenvalue_f_dependence}, the dependence of eigenmodes on $\tilde{f}$ is shown.
The dependence becomes weaker as the mode number $n$ or the azimuthal quantum number $m$
increase. For $\tilde{f}\sim {\cal O}[10^{-1}]$, the eigefrequency may be regarded not to depend on $\tilde{f}$.
The stellar frequency $\nu_\star$(Hz) is related to $\tilde{f}$ as
\beq
	\tilde{f} = 8.07\times 10^{-5} \nu_\star \qty(\frac{h_{out}}{100\mbox{m}})^\half \gamma.
\eeq
Assuming an extreme case of $\nu_\star\sim 1$kHz, the aspect ratio $\gamma\sim 10$, we have $\tilde{f}\sim 0.8$.
Therefore, we expect $\tilde{f}$ to be much less than ${\cal O}[1]$ for a slowly rotating star,
as far as the depth of the ocean is very small compared with the radius of the magnetic pole. We may then neglect
a dependence of eigenfrequencies on $\tilde{f}$.

\begin{figure}[h]
\includegraphics[width=9cm]{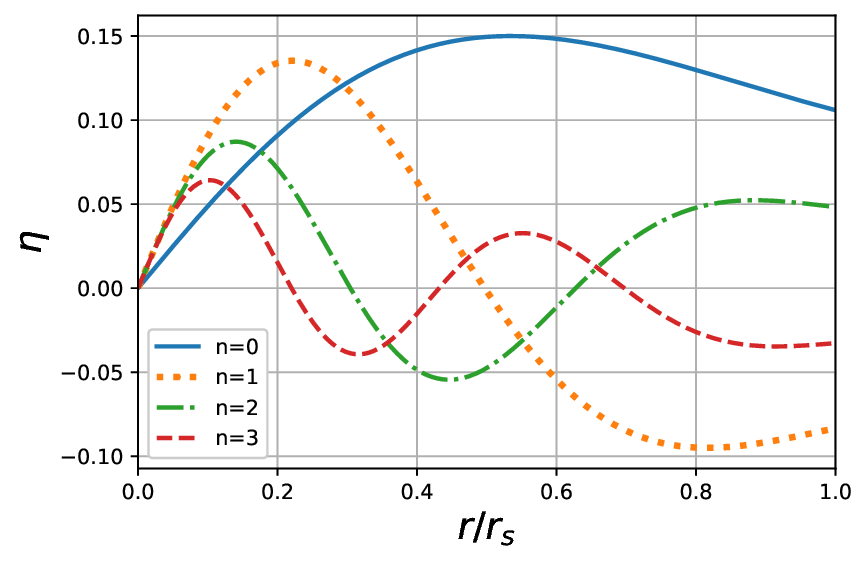}
\caption{\label{fig: eigenfunction m1 case} Eigenfunction for the lowest four eigenmodes of $m=1$.
$n$ is the mode number (the number of nodes except the origin). The model parameters are $(\alpha, \tilde{f}) = (0.9, 0.0807)$.}
\end{figure}

In Fig.\ref{fig: eigenfunction m1 case}, the lowest order four eigenfunctions of $m=1$ are plotted.
The number of nodes (order $n$) is counted excluding the origin. The region plotted is inside
the magnetic pole. Outside the region plotted, the eigenfunctions are monotonically (exponentially) damped.

\begin{figure}[h]
\includegraphics[width=9cm]{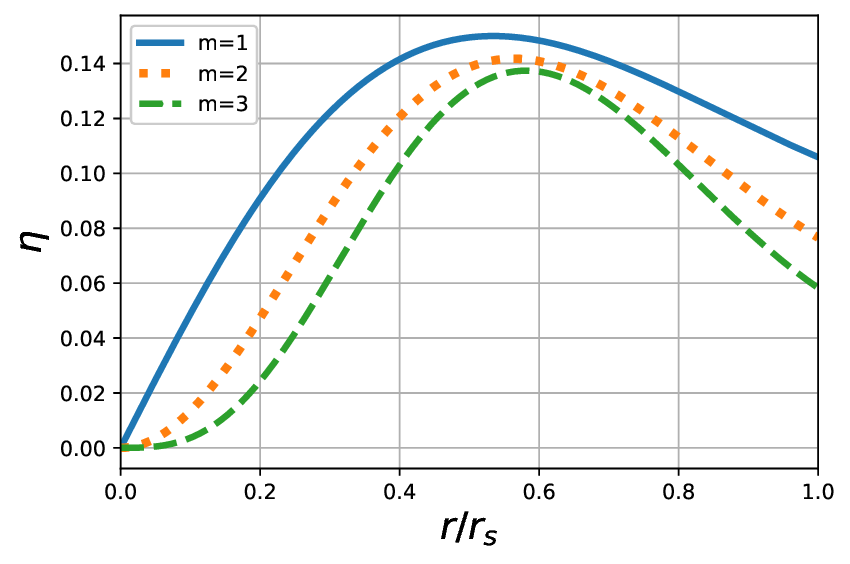}
\caption{\label{fig: eigenfunction n0 case} The fundamental eigenmode for $m=1, 2, 3$.
The model parameters are $(\alpha, \tilde{f}) = (0.9, 0.0807)$.}
\end{figure}

The difference of the fundamental mode with different $m$ is shown in Fig.\ref{fig: eigenfunction n0 case}.
We note that $\eta \sim r^m$ as $r\to 0$.

\begin{figure}[h]
\includegraphics[width=9cm]{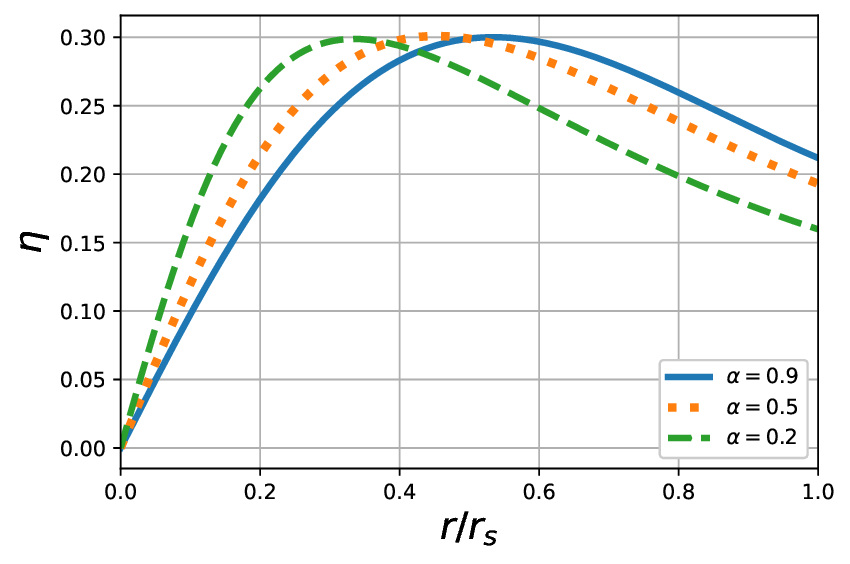}
\caption{\label{fig: eigenfunction alpha case} The fundamental eigenmode for $m=1$
with different values of $\alpha$. $\tilde{f} = 0.0807$.}
\end{figure}

The dependence of the fundamental $m=1$ eigenfunction on $\alpha$ is exhibited in Fig.\ref{fig: eigenfunction alpha case}.
The distribution of the mode amplitude shifts towards the origin as $\alpha$ decreases, i.e., as the field strength in the
magnetic pole increases.

\begin{figure}[h]
\includegraphics[width=9cm]{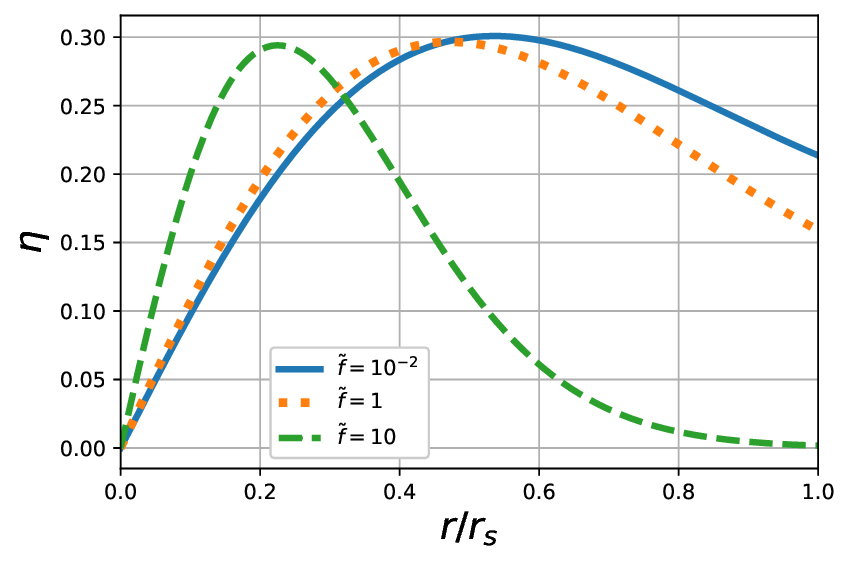}
\caption{\label{fig: eigenfunction f case} The fundamental eigenmode for $m=1$
with different values of $\tilde{f}$. $\alpha=0.9$.}
\end{figure}

Figure \ref{fig: eigenfunction f case} shows the profile of the $m=1$ fundamental mode for the $\alpha=0.9$ case.
As $\tilde{f}$ increases, the peak position of the amplitude moves towards the origin. At the same time, the full-width-half-maximum of the eigenfunction decreases, thus the oscillation is more strongly trapped in the magnetic pole.
Given the magnetic field strength of the pole, the faster stellar rotation, parametrized by the larger $\tilde{f}$, 
is favorable for the trapping of the eigenmode.

\section{Discussion : Implication to low-frequency QPOs in neutron star X-ray binaries\label{sec: discussion}}
The dimensionless eigenfrequency is converted to a physical value by accounting for two relativistic factors.
One is that the general relativistic effect modifies the gravitational acceleration seen by a local static observer
on the stellar surface.
The Newtonian expression of the surface gravitational
acceleration is $g_0 = GM/R^2$, but the acceleration measured by the static observer
at the stellar surface $r=R$ is $g = g_0/\sqrt{1-2GM/c^2R}$. Secondly, the eigenfrequency
is the value locally defined by the static observer's proper time. Therefore, the redshift
factor $\sqrt{1-2GM/c^2R}$ needs to be multiplied to obtain the frequency seen by an observer
on the Earth. 

X-ray observations of accreting neutron stars with low-/high-mass  companions
have revealed that these systems show different types of characteristic flux variations called
quasi-periodic oscillations (QPOs) \cite{vanderKlis1989, vanderKlis2004}. For low-mass systems, the neutron star
may spin at a frequency close to a few tens to a few hundred Hz range, and its modulation on accreting matter may
result in the QPOs. There are a variety of models to explain the nature of these QPOs 
by the beat of the rotational frequency of the star and the orbital frequency of the accretion disk (see the review paper
\cite{vanderKlis2004} for those models). Other types of models look at the oscillation of accretion
flow in the magnetic polar caps, i.e., photon bubble oscillation model \cite{Klein1996}; moving hot spot
model \cite{Bachetti_etal2010}; 
radiation-MHD oscillation model of accretion column \cite{Zhang_etal2023};
magnetospheric oscillation model \cite{Shi_etal2014}.

Low-frequency QPOs in sub-Hertz ($1-10^3$mHz) frequency range are also observed
in low-/high-mass systems \cite{Manikantan_etal2024, Li_etal2024, Malacaria_etal2024, Yang_Wang2025}. A plausible model of these QPOs needs
to explain a much lower frequency than the dynamical one at the surface of a star ($\sim 1$kHz).
For LMXBs, \cite{Heger_etal2007} introduces nonlinear pulsations of marginally stable nuclear burning 
at the surface of an accreting star, whose thermal timescale may be $1$ks.



The observed low-frequency QPOs in high-mass X-ray binaries (HMXBs) are not explained by the nuclear burning instability, and there is 
quite a diversity in their modeling (see \cite{Yang_Wang2025}). One of the models invokes the Keplerian orbit
of a blob of gas at the inner edge of the accretion disk \cite{Van_der_Klis1987}. A similar mechanism, but considering
the sidebands of the neutron star's spin frequency due to the orbiting blob, is introduced by \cite{Kommers1998}.
Beat of the orbital frequency at the inner edge of the disk with the stellar spin frequency has also been considered \cite{Alpar_Shaham1985}.
An instability of accretion column flow onto a magnetic pole is invoked to explain a low-frequency QPO at a
rather high accretion rate \cite{Li_etal2024}. The intermittent accretion from the corotation radius in the accretion disk
onto the central star is proposed to explain another low-frequency QPO \cite{Li_etal2024}. The inner accretion disk
may be unstable to non-axisymmetric pattern formation (polygonal vortices), and it may result in some of the QPOs \cite{Ding_etal2021}.
Finally, the stellar magnetic field threaded in the inner accretion disk may warp it and induce precession of
the disk, which leads to low-frequency QPOs \cite{Shirakawa_Lai2002}. In contrast to the low-frequency QPOs in
LMXBs, those in HMXBs may not have a unique explanation applicable to all the sources, and seemingly
QPOs with different origins may coexist in the same system.

Based on the trapped surface gravity wave, we propose another simple alternative interpretation of low-frequency QPOs 
in the X-ray binary systems whose accretor is a strongly magnetized neutron star.
The accreted matter from the companion star (fed by its wind) is channeled to the magnetic pole
and forms an accretion column. The accretion column is expected to have a hydrostatic
layer below the accretion shock \cite{Baan-Treves1973, Davidson1973, Basko-Sunyaev1976}.
Highly non-isotropic radiation perpendicular to the magnetic column enables the very
large apparent luminosity, which may be close to or exceed the Eddington limit of a typical neutron star.
When the surface gravity wave is trapped at the bottom of the hydrostatic column, its stratification
is modulated at the frequency. The radiation from the column may be naturally modulated by the low-frequency
of the surface gravity wave and leads to mHz QPOs. 
When the accretion rate is sufficiently high, the column may be unstable to the formation of
photon bubbles, which are regions of low-density and radiation-dominated plasma \cite{Arons1992, Gammie1998, Begelman2006}.
Oscillations of photon bubbles may be responsible for some of the QPOs in the kHz range \cite{Klein1996}, but
are too high in frequency to explain the mHz QPOs. Nevertheless, the bubble oscillations and their radiations will be modulated
by the surface gravity wave, since the lower boundary of the column is modulated by the low-frequency
trapped waves.

\begin{figure}[h]
\includegraphics[width=9cm]{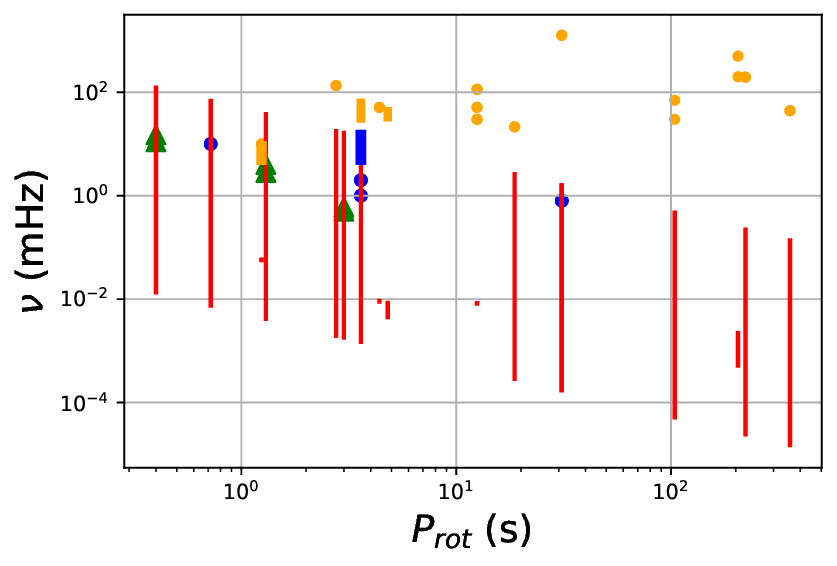}
\caption{\label{fig: comparison with HMXB} Comparison of the eigenfrequency of $m=1$ fundamental mode
with the observed low-frequency QPOs in HMXBs and PULXs. 
HMXBs' data are taken from \cite{Yang_Wang2025} and \cite{Salganik2026}, 
while those for PULXs are taken from \cite{Cemeljic2025}. The horizontal axis is the spin period of neutron stars,
while the vertical axis is the QPO frequency in mHz. The blue dots and lines correspond to the observed
QPO frequency that falls in the range of the fundamental frequency. The orange dots are observed QPOs
that do not match the fundamental frequency.
The red lines show the eigenfrequency of $m=1$ fundamental mode for the magnetic field strength
spanning from $B=10^{12}$G (lower ends) to $B=10^{14}$G (upper ends), except
for the HMXB systems for which the magnetic field strength is inferred by the cyclotron resonance scattering features 
(Her X-1\cite{Gruber2021};Cen X-3\cite{Liu2022};IGR J19294+1816\cite{Tsygankov2019};V 0332+53\cite{Makishima1990,DAi2025}; RXJ0440.0+4431\cite{Li2024, Epili_Wang2025}). For these systems, the magnetic field inferred from the
observations is adopted. To compute the model frequency,
$h_{out}=10^2$m is assumed. 
The systems are, from the left, NGC7793 P13, SMC X-1, Her X-1, M82 X-2, 4U 1901+03, M51 ULX7,  4U 0115+63, V 0332+53,
Cen X-3, IGR J19294+1816, KS 1947+300, XTE J10111.2-7317, 1A 05335+262, RX J0440.9+4431, XTE 1958+034, SAX J2103.5+4545. 
The green triangles mark the QPOs in the ULX sources.}
\end{figure}

For HMXB systems where low-frequency QPOs are observed, the fundamental $m=1$ modes are
compared in Fig.\ref{fig: comparison with HMXB}. It should be noted that the Coriolis factor is computed
by assuming the cosine of the polar angle to be $1/\sqrt{3}$. The aspect ratio $\gamma$ of the radius of the pole
to the ocean depth of the unmagnetized region is fixed as unity. 
Also, the magnetic field is computed by assuming $h_{out}=10^2$m (see Eq.(\ref{eq: B-field estimation}))
and all the stars are assumed to have $M=1.4M_\odot$ and $R=10^6$km. The model lines for each star (red solid) are for $10^{12}\le B(\mbox{G})\le 10^{14}$,
where $B=10^{12}$G corresponds to the lower end, while $B=10^{14}$G to the upper end.
As is seen in the figure, for the spin period and the strength of the magnetic field considered here, 
the QPOs observed in the stars with $P_{rot}\ge 10$s
are not explained by the $m=1$ fundamental mode. Our model
may apply to the low-frequency QPOs found in fast spinning systems. 
It should also be noted that the sub-mHz trapped modes in the lower magnetic
field stars, if they exist, may have been hard to detect in those observations (see e.g., the power-spectrum density plots in \cite{Manikantan_etal2024}).
Since eigenmodes with higher $m$ or $n$ have lower frequencies, these overtones do not explain the observed 
QPOs for $P_{rot}\ge 10$s either.
For the stars with faster rotation, the $m=1$ fundamental eigenmode may be consistent with some of the QPOs.

In the figure, QPOs observed in three of the ultra-luminous X-ray sources (ULXs) are plotted as well. 
ULXs are X-ray sources whose luminosity satisfies $L>10^{39}{\rm erg}{\rm s}^{-1}$ \cite{King2023}. For the higher end, the luminosity may reach $10^{41}-10^{42}{\rm erg}{\rm s}^{-1}$ and may exceed the Eddington luminosity of a neutron star or a usual stellar mass black hole.
Although the high luminosity exceeding the Eddington limit may be simply explained by such massive 
accretors as intermediate mass black holes, there are observational clues that not all of them are
more massive than the typical stellar mass black holes or neutron stars.
Especially, among more than 1000 candidates of ULXs found, fewer than a dozen show periodic pulsations, which are interpreted as
the spin period of accreting neutron stars (pulsating ULXs, PULXs). Currently, three PULXs show low-frequency QPOs.
\cite{Veresvarska2025} and \cite{Cemeljic2025} apply the magnetic precession model of accretion disk 
by \cite{Pfeiffer-Lai2004} to explain the QPOs.
The model explains that the QPO frequencies in PULXs are proportional to the spin frequency of the accretor.
Also in our model, this proportionality is naturally explained by the scaling of frequency (Eq.(\ref{eq: scaling of frequency})).
As is seen in Fig.\ref{fig: comparison with HMXB}, the model frequencies may be consistent with the observations, if the magnetic
field of the accretor in these PULXs is higher than assumed in \cite{Cemeljic2025} ($B\le {\cal O}[10^{11}]$G), which is based on the super-critical accretion
disk model of \cite{KLK2017}. 

\section{Summary and Conclusion}
We consider the surface gravity waves on a neutron star ocean when magnetic poles are present.
As the interaction between the tidal waves and the coastal geography of an island enables 
the existence of trapped modes in the terrestrial ocean, the magnetic field may also trap 
the surface gravity waves to form a discrete spectrum of eigenmodes
in the neutron star ocean.
We have shown that there exist eigenmodes of trapped waves whose spectrum is reminiscent
of that of g-modes in asteroseismology.

The trapped waves in the magnetic pole modulate the emission from the accretion column of gas above,
and we may expect the low-frequency QPOs observed in the accreting neutron stars.
The estimate here shows that the QPO frequencies for stars in PULXs and in HMXBs with $P_{rot}\le 10$s are consistent
with the model, although the eigenfrequency for the stars with slower rotation is much lower than the QPOs observed. 
This does not necessarily mean that the trapped surface gravity wave
modes do not exist. 
In view of the observed frequency spectra of the light curve of X-ray binaries 
(e.g,, \cite{Li_etal2024}), the sub-mHz region of the frequency has not been fully investigated thus far
for most of the HMXB systems. It would be
interesting to expand the observational study of QPOs to the sub-mHz range to look for the trapped waves at the magnetic
pole.
\begin{acknowledgements}
The data that support the findings of this article are openly available. See \url{https://github.com/shin-yoshida-astrophys/surface-gravity-wave}.
\end{acknowledgements}

\appendix*

\section{Shallow water wave equations\label{app: shallow water}}
Linearizing Eq.(\ref{eq: momentum}) and Eq.(\ref{eq: induction}), we have,
\beq
	\pdv{t}\delta v^x = -f\delta v^y -\frac{1}{\rho}\pdv{\delta p}{x} + \frac{1}{4\pi\rho}\qty(B^z\pdv{z}\delta B_x - \pdv{x}(B^z\delta B_z)),
\eeq
and
\beq
	\pdv{t}\delta v^y = f\delta v^x -\frac{1}{\rho}\pdv{\delta p}{y} + \frac{1}{4\pi\rho}\qty(B^z\pdv{z}\delta B_y - \pdv{y}(B^z\delta B_z)),
\eeq
where $v^x, v^y, v^z$ are Cartesian components of fluid velocity. 

Our interest is in a generalization of shallow water wave theory \cite{Pedlosky1987book}
in the case of variable sea level around a magnetic pole. Firstly, as is the case with the original shallow water theory,
we here adopt the planar ansatz for the perturbed velocity,
\beq
	\pdv{\delta v^x}{z} = 0 = \pdv{\delta v^y}{z}.
	\label{eq: planar condition}
\eeq
From x- and y-components of the magnetic induction equation (Eq.\ref{eq: induction}), we have
\begin{align}
	\pdv{\delta B^x}{t} = B^z\pdv{\delta v^x}{z}\\
	\pdv{\delta B^y}{t} = B^z\pdv{\delta v^y}{z},
\end{align}
thus,
\beq
	\delta B^x = 0 = \delta B^y.
	\label{eq: perturbed tangential B}
\eeq
Then, from the solenoidal nature of magnetic field, we have
\beq
	\pdv{\delta B^z}{z} = 0.
	\label{eq: perturbed normal B}
\eeq
It leads to the tangential components of the perturbed equation of motion,
\begin{align}
	\pdv{\delta v^x}{t} = -f\delta v^y -\pdv{x}\qty(\frac{\delta p}{\rho} + \frac{B^z\delta B_z}{4\pi\rho})\\
	 \pdv{\delta v^y}{t} = f\delta v^x -\pdv{y}\qty(\frac{\delta p}{\rho} + \frac{B^z\delta B_z}{4\pi\rho}),
\end{align}
whereas the normal component is,
\beq
	\pdv{\delta v^z}{t} = -\pdv{z}\frac{\delta p}{\rho}.
\eeq
It should be noticed that the gravitational acceleration $g$ does not appear in the perturbed momentum equation
in z-direction, because the gravity term in Eq.(\ref{eq: momentum}) is balanced by the unperturbed pressure gradient.
We then introduce $\delta P = \delta p+ B^z \delta B^z/4\pi$. By using Eq.(\ref{eq: perturbed normal B}),
we may write all the components of the equation of motion as,
\begin{align}
	\pdv{\delta v^x}{t} = -f\delta v^y -\pdv{x}\frac{\delta P}{\rho},\label{eq: perturbed momentum x}\\
	\pdv{\delta v^y}{t} = f\delta v^x -\pdv{y}\frac{\delta P}{\rho}\label{eq: perturbed momentum y}
\end{align}
and 
\beq
	\pdv{\delta v^z}{t} = -\pdv{z}\frac{\delta P}{\rho},
	\label{eq: vertical momentum}
\eeq
since the magnetic term does not depend on z-coordinate.
In the shallow water approximation, where the characteristic depth of the fluid $D$ is much smaller than the
characteristic length scale of the waves in horizontal direction $L$, i.e., $D/L\equiv \epsilon\ll 1$,
the left hand side of Eq(\ref{eq: vertical momentum}) is ${\cal O}[\epsilon^2]$ times the right hand side
\cite{Pedlosky1987book}. Thus we have
\beq
	\pdv{z}\frac{\delta P}{\rho} = 0,
\eeq
i.e., the hydrostatic balance in z-direction for the perturbed state. It is integrated to define a function $\eta(x, y, t)$,
\beq
	\frac{\delta P}{\rho} = g\eta(x, y, t).
	\label{eq: pressure perturbation in eta}
\eeq
In fact, $\eta$ defines the displaced surface of the ocean as
\beq
	h(x, y, t) = \bar{h}(x, y) + \eta(x, y, t).
\eeq
Physically, Eq.(\ref{eq: pressure perturbation in eta}) states that the incremental height
$\eta$ to the original surface $\bar{h}$ increases the pressure below by the same amount
of the added weight of the fluid.

Next, by integrating the equation of continuity
\beq
	\pdv{v^x}{x} + \pdv{v^y}{y} + \pdv{v^z}{z} = 0
\eeq
in z-direction from $0$ to $\bar{h}+\eta$, we obtain
\beq
	(\bar{h} + \eta) \qty(\pdv{v^x}{x} + \pdv{v^y}{y}) + \qty[v^z]_0^{\bar{h}+\eta} = 0.
	\label{eq: continuity total}
\eeq
The kinetic boundary condition at the surface of the ocean is that the vertical velocity
is equal to the Lagrangian derivative of the surface profile $h = \bar{h} + \eta$,
\beq
	v^z = \dv{h}{t},
\eeq
where $h(x, y, z, t)$ is the instantaneous surface profile of the ocean.
Using this relation to Eq.(\ref{eq: continuity total}) in the linear order of perturbation, 
we have the equation for $\eta$
\beq
	\pdv{\eta}{t} + \pdv{x}\qty[\bar{h}\delta v^x] + \pdv{y}\qty[\bar{h} \delta v^y] = 0.
	\label{eq: perturbed continuity}
\eeq

Eq.(\ref{eq: perturbed momentum x}), (\ref{eq: perturbed momentum y}), (\ref{eq: pressure perturbation in eta})
and (\ref{eq: perturbed continuity}) constitute the basic equations of the linearized surface wave.
It should be noted that the planar ansatz of perturbed velocity from which we start reducing the system of equations
is consistent with these basic equations, thus the circle closes.

The horizontal components of velocity perturbation are obtained by solving Eqs.(\ref{eq: perturbed momentum x}) and (\ref{eq: perturbed momentum y}) for given eigenfrequency and eigenfunction $\eta$.
Perturbed magnetic field has only z-component $\delta B^z$ and follows the perturbed induction equation,
\beq
	\pdv{\delta B^z}{t} = -\delta v^x\pdv{B^z}{x} - \delta v^y\pdv{B^z}{y}.
	\label{eq: delta Bz}
\eeq

\begin{figure}[h]
\includegraphics[width=9cm]{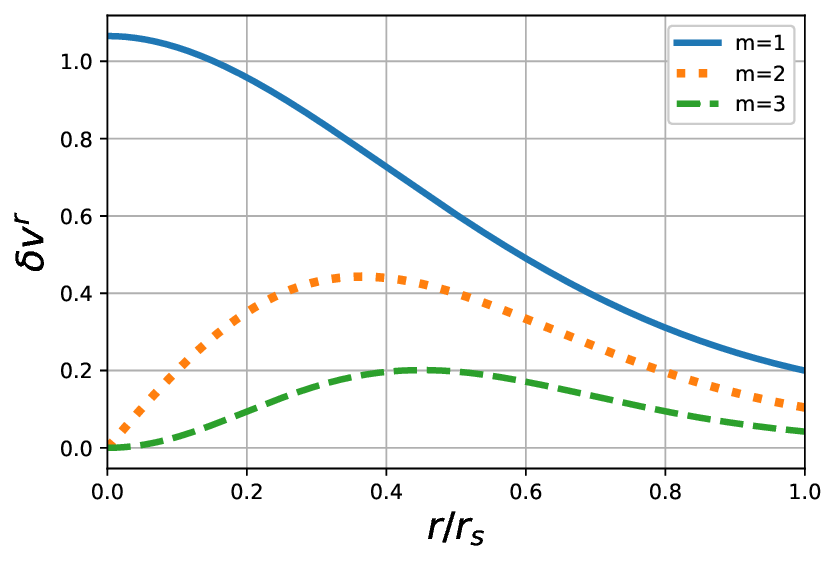}
\caption{\label{fig: eigenfunction of vr} Radial velocity perturbation of fundamental modes. The equilibrium parameters are the sama as Fig.\ref{fig: spectrum}.}
\end{figure}

\begin{figure}[h]
\includegraphics[width=9cm]{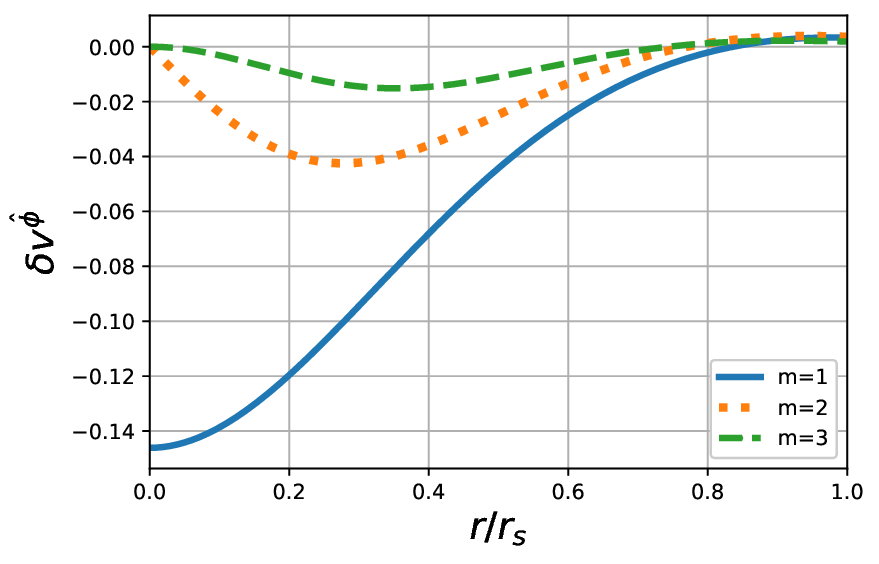}
\caption{\label{fig: eigenfunction of vphi} Azimuthal velocity perturbation of fundamental modes. The equilibrium parameters are the sama as Fig.\ref{fig: spectrum}.}
\end{figure}

In Fig.\ref{fig: eigenfunction of vr} and \ref{fig: eigenfunction of vphi}, velocity perturbation for cylindrical $r$ and $\phi$ components as a function of $r$ 
are plotted for the fundamental modes with $m=1, 2, 3$.
It should be remarked that the $m=1$ velocity perturbations are finite at the origin, while they
are zero for $m\ge 2$. For $m=1$, the oscillation is dipolar and the regularity condition
of the mode is that the radial derivatives of the perturbed velocity vanish there. 
For $m\ge 2$, the regularity at the origin enforces that the perturbed velocity components vanish.

\begin{figure}[h]
\includegraphics[width=9cm]{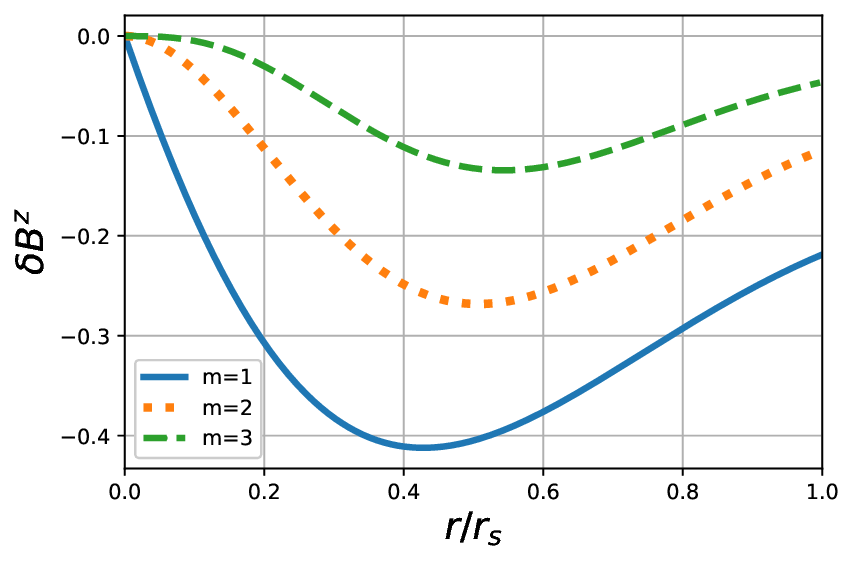}
\caption{\label{fig: eigenfunction of Bz} Perturbed magnetic field (z-component) of fundamental modes. The equilibrium parameters are the sama as Fig.\ref{fig: spectrum}.}
\end{figure}

In Fig.\ref{fig: eigenfunction of Bz}, the perturbed magnetic field profile is shown. As is seen from
Eq.(\ref{eq: delta Bz}), the phase of the magnetic perturbation $\delta B^z$ and that of the velocity
perturbation differ by $\pi/2$. 
\bibliography{sww}

\end{document}